\documentclass[%
aps,
prd,
amsmath,amssymb,
superscriptaddress,
12pt]{revtex4-2}

\usepackage[T1]{fontenc}
\usepackage[utf8]{inputenc}
\usepackage{amsmath,amsthm,amssymb,amsfonts,enumerate,enumitem,graphicx,xcolor,multirow,soul}
\usepackage[normalem]{ulem}
\usepackage[pdfusetitle,bookmarksopen]{hyperref}
\usepackage{cleveref}

\crefname{equation}{eq.}{eqs.}

\newcommand{\bra}[1]{\langle #1 |}
\newcommand{\ket}[1]{| #1 \rangle}

\def\ba#1\ea{\begin{align}#1\end{align}}
\newcommand{\nn}{\nonumber}
\newcommand{\ra}{\rangle}

\newcommand\uconn{Physics Department, University of Connecticut, Storrs, CT 06269-3046, USA}

\renewcommand{\hl}[1]{#1}
\renewcommand{\sout}[1]{}

\begin{document}
	
\title{Using lattice chiral effective theory to study $\pi\pi$ scattering}
\author{Cameron Cianci}
\affiliation{\uconn}
\author{Luchang Jin}
\affiliation{\uconn}
\author{Joshua Swaim}
\email{joshuatylerswaim@gmail.com}
\affiliation{\uconn}
\date{\today}

\begin{abstract}
We use lattice field theory to study the finite-volume energy spectrum of the $\pi\pi$ system in $SU(2)$ chiral effective field theory (ChEFT) at leading order in the chiral expansion. \hl{This finite-volume spectrum can be directly related to the (infinite-volume) $\pi\pi$ scattering phase shifts by L\"uscher's formula.} We compare our results to the finite-volume spectrum obtained from lattice QCD \hl{by the RBC-UKQCD collaboration}. Our calculation and the lattice QCD calculation are both performed with the physical pion mass and the same \sout{physical volume}\hl{lattice volume (as measured in physical units)}. However, we find significant differences between the two calculations in the isospin $I=0$ channel. In particular, there is a nearly stable $\sigma$ resonance in our lattice ChEFT calculation, which is absent in the lattice QCD calculation. This likely indicates that ChEFT does not converge well with a naive lattice regularization.
\end{abstract}

\maketitle

\section{Introduction}
Chiral effective field theory (ChEFT) was introduced by Weinberg in 1979 as a low energy effective theory for QCD \cite{Wein_1979}. While \sout{chiral \textit{perturbation} theory (ChPT) has proven a very powerful theoretical tool}\hl{ChEFT is often employed in conjunction with perturbation theory}, there are good reasons for studying ChEFT non-perturbatively\sout{using lattice Monte Carlo}. \hl{For example, we can use chiral symmetry through the PCAC relations to obtain many useful results}~\cite{Scherer:2012xha}. \hl{We can get even more non-perturbative information using lattice Monte Carlo. There are many reasons to be interested in doing this.}

\begin{itemize}
	\item First, lattice ChEFT could aid lattice QCD calculations. \sout{ChPT}\hl{Perturbative ChEFT} is used extensively to perform various extrapolations on lattice QCD results (for example, extrapolations to infinite volume or to physical pion mass; see Ref.~\cite{Shanahan_review} for a review). These extrapolations could also be performed using lattice ChEFT, which would allow direct comparison of the observables in the effective theory with QCD, even at the level of correlation functions.
    \item Second, by matching lattice ChEFT with lattice QCD, we could determine the ChEFT low energy constants in the lattice regularization, and, with some additional calculation, in perturbative renormalization schemes such as $\overline{\text{MS}}$. 
    \item Third, lattice ChEFT (involving only pion degrees of freedom) could provide useful insights for more complicated lattice simulations of nuclear ChEFT. In nuclear physics, lattice ChEFT has long been used to study the spectrum, structure, and scattering of atomic nuclei~\cite{Lee_power_count_2004, Borasoy_LO_2006, Borasoy_NLO_2007, Epelbaum_dilute_neutron_ground_state_2009, Epelbaum_three_body_int_2009, Epelbaum_nuclei_ground_state_2010, Epelbaum_Hoyle_state_2011, Epelbaum_triple_alpha_2013, Epelbaum_spec_16O_2013, Lahde_med_mass_nuclei_2013, Epelbaum_NNNLO_2014, Lahde_sym_sign_method_2015, Elhatisari_alpha_scattering_2015, Elhatisari_nucleon_deuteron_scattering_2016, Elhatisari_nuc_binding_near_phase_trans_2016, Elhatisari_nuclear_clustering_2017, Alarcon_neutron_proton_scattering_2017, Li_neutron_proton_NNNLO_2018, Klein_Symanzik_improvements_2018, Klein_Tjon_band_2018, Madeira_QMC_dyn_pions_2018, Ahmadi_new_reg_2020, Zhang_DDK_system_2024, Wu_charged_nucleon_int_2025, Liu_QMC_2025}. It should be noted that most of these calculations use instantaneous pions (Ref.~\cite{Lee_power_count_2004} is one of the exceptions). See Ref.~\cite{Madeira_QMC_dyn_pions_2018} for a more recent proposal to introduce dynamical pions using the Quantum Monte Carlo method.
    \item Fourth, a more direct mapping between chiral effective theory and QCD could add insight into chiral effective theory. For example, it could contribute to understanding the nature of the $\sigma$ resonance (see Ref.~\cite{Li_sigma_review_2025} for a review discussing recent progress on this question).

\end{itemize}

The leading-order ($\mathcal{O}(p^2/\Lambda^2)$) $SU(2)$ chiral effective theory is equivalent to
the $O(4)$ non-linear sigma model, which in turn is the strongly-interacting limit of the $O(4)$ linear sigma model~\cite{Gell-Mann_1960}.
The $O(4)$ (non-)linear sigma model has been frequently studied on the lattice with Monte Carlo calculations.
It is often used as a ``toy'' model to test various algorithmic and theoretical techniques, such as L\"uscher's formula~\cite{Nishimura_Luscher_test_1992,Gockeler_Luscher_test_1994} and the maximum entropy method \cite{Yamazaki_max_entropy_test_2002}. It has also been employed to study the Higgs sector of the standard model~\cite{Hasenfratz_Higgs_bound_1988,Gockeler_Higgs_bound_1993}.
In this work, we view it as the chiral effective theory at leading order in the effective field theory expansion.
We choose its parameters to give the pions their physical mass and decay constant.
To our knowledge, this is the first time that lattice ChEFT is used to study \sout{the physical}$\pi\pi$\sout{spectrum} \hl{scattering and compared with first-principles lattice QCD calculations of the same process. The finite-volume $\pi\pi$ energy spectrum is directly related to (infinite-volume) scattering phase shifts by L\"uscher's formula \cite{Luscher:1990ux}, and the lattice QCD calculations agree well with experimental measurements \cite{RBC_pipi_scattering}}.

We find that our lattice simulation  of the sigma model does \emph{not} reproduce the low-energy spectrum of QCD.
This is likely because the lattice regularization introduces power divergences which \hl{enhance higher order terms in}\sout{disrupt} the normal power-counting for chiral effective theory (see \Cref{power_counting}). This problem might be solved by using a more sophisticated lattice regularization (for example, by introducing smearing).

In \Cref{LSM}, we introduce the linear sigma model and discuss its connection to $SU(2)$ chiral effective theory. In \Cref{results}, we present the results of our calculations. We show that we are indeed simulating in the non-linear sigma model regime. We then show how various observables depend on our model parameters, demonstrating that the physical point is unique, and that it lies near a phase transition. Finally, we compare the energy spectrum of our theory to lattice QCD calculations, noting significant disagreement in the isospin $I=0$ channel. Because the physical point is unique, the low energy constants
cannot be adjusted to fix this discrepancy. In \Cref{power_counting}, we discuss the likely reason for this discrepancy and a possible resolution.

\section{Chiral Effective Theory and The Linear Sigma Model} \label{LSM}

\subsection{The Linear Sigma Model}

In principle, it is possible to construct an effective scalar field theory in terms of the pion fields which exactly reproduces the dynamics of QCD. In \Cref{sec:exact-effective-theory-for-qcd}, we show this is true by explicitly constructing such an effective action. The exponential of the resulting action is real, but it is not guaranteed to always be positive.
While this construction is explicit \hl{and proves that a sufficiently sophisticated effective action can describe QCD to arbitrary high accuracy,} it is not yet practical \sout{for either numerical or perturbative calculations}\hl{to obtain this sophisticated action in a form that is suitable for effective theory calculations. 
In the following numerical work, we instead approximate the effective action with the leading order terms in the conventional chiral expansion, with the understanding that more accurate results can be obtained if more higher-order terms are included in the effective action.}

In the standard treatment, ``one writes down the most general
possible Lagrangian, including all terms consistent with assumed symmetry principles'' \cite{Wein_1979}.
The resulting Lagrangian is in general non-local. However, we can expand this non-local function in terms of derivatives of the scalar fields.
For a process involving energy scale $E$, most terms contribute corrections which are suppressed by powers of $E / \Lambda$, where $\Lambda$ represents the cutoff scale of the effective theory. If we collect the pion degrees of freedom into an SU$(2)$ field $U$, the leading-order SU$(2)$ chiral effective Lagrangian (after Wick rotation to Euclidean time) is
\ba \mathcal{L}_{\text{SU}(2)}=\frac{F_0^2}{4}\text{Tr}\left[\sum_\mu(\partial_\mu U)\partial_\mu U^\dagger\right]+\frac{F_0^2B_0M}{2}\text{Tr}(U^\dagger+U), \ea 
where $F_0$, $B_0$, and $M$ are scalar parameters. This is also called the SU$(2)$ non-linear sigma model.
Because SU$(2)$ is diffeomorphic to the 3-sphere, we can replace our dynamical variable $U$ with a set of four fields $\phi_i$ constrained so that the sum of their squares is constant. A convenient diffeomorphism is given by
\ba U=\frac{\phi_0}{F_0}+i\sum_{j=1,2,3}\frac{\phi_j}{F_0} \tau_j,\ea 
where $\tau_j$ are the Pauli sigma matrices, and $\sum_{j=0}^3 \phi_j^2=F_0^2$. In terms of these new dynamical variables, 
\ba
\mathcal{L}=\frac{1}{2}\sum_\mu\sum_{j=0}^3(\partial_\mu\phi_j)^2+\alpha\phi_0~\text{, with }\alpha\equiv 2F_0B_0M\text{ and constraint }\sum_{j=0}^3 \phi_j^2=F_0^2.
\ea 

There is a closely related model in which the four fields are no longer exactly constrained to lie on the surface of a 3-sphere: the $O(4)$ linear sigma model 
\ba \mathcal{L} = \frac{1}{2}\sum_\mu\sum_{i=0}^3(\partial_\mu\phi_i)^2+\frac{m^2}{2}\sum_{i=0}^3\phi_i^2+\frac{\lambda}{4!}\left(\sum_{i=0}^3 \phi_i^2\right)^2+\alpha \phi_0.\ea 
The $O(4)$ symmetry of this model can be broken both spontaneously (for $m^2<0$) and explicitly by the $\alpha \phi_0(x)$ term. In the symmetry-broken phase, the fields are dynamically constrained to lie close to the surface of a 3-sphere, with the strength of the constraint controlled by the parameter $\lambda$.
\hl{Unlike the non-linear sigma model, the linear sigma model is renormalizable. However, as L\"uscher demonstrated~\cite{Luscher:1987ay, Luscher:1988uq},
the theory likely does not have a non-trivial continuum limit. Of course, this does not affect our use of the linear sigma model as an effective theory, which has an explicit cutoff.}
In the limit as $\lambda\to \infty$ (with $\frac{m^2}{\lambda}$ held fixed), we return to the non-linear sigma model.
In this paper, we study the linear sigma model numerically with very large $\lambda$ ($\lambda=10^4$). To discretize the theory for the lattice, we use the Lagrangian density
\ba 
\mathcal{L}(x)=\frac{1}{2}\sum_{\mu}\sum_{i=0}^3 \big(\phi_i(x+\mu)-\phi_i(x)\big)^2+\frac{m^2}{2}\sum_{i=0}^3\phi_i(x)^2+\frac{\lambda}{4!}\left(\sum_{i=0}^3\phi_i(x)^2\right)^2+\alpha\phi_0(x).\ea 

\subsection{Particle Identifications}\label{sec_particle_ids}
In our theory, \hl{when there is no explicit symmetry breaking ($\alpha=0$), the $O(4)$ symmetry can be spontaneously broken. The symmetry can break along any direction, described by some linear combination of the four fields. We may simply redefine the fields so that the direction of symmetry breaking lies along $\phi_0$. Then $\phi_0$ will have a non-zero vacuum expectation value, while all the other three fields will still have a vacuum expectation value of 0, respecting a reduced $O(3)$ symmetry. These three fields correspond to massless Goldstone bosons. When we add in explicit symmetry breaking ($\alpha\neq 0$),} the explicit symmetry-breaking term $\alpha\phi_0$ ensures that the $O(4)$ symmetry will be broken in such a way that \hl{$\phi_0$ takes on the non-zero vacuum expectation value without the need for redefinition. This means the fields} $\phi_1$, $\phi_2$, and $\phi_3$ correspond to the pseudo-Goldstone bosons. \hl{As long as $\alpha$ is small, their masses will be small.} Therefore, these fields are identified as pions. The remaining degree of freedom is the sigma field $\sigma=\phi_0-\langle\phi_0\rangle$, where $\langle \phi_0\rangle$ is the vacuum expectation value of $\phi_0$. The sigma degree of freedom disappears in the $\lambda\to\infty$ limit, although the theory will still contain an associated resonance.

\subsection{Pion Decay Constant}\label{sec_pion_decay_constant}
The pion decay constant $F_\pi$ may be defined in the continuum by the matrix element of the axial current between the vacuum and a single-pion state
\ba \bra{0}A^i_\mu(x,t) \ket{\pi(p)}=F_\pi p^\mu e^{-p_0 t-i\mathbf{p}\cdot\mathbf{x}}.\ea 
Here we define $A_{\mu,i} \equiv \bar\psi\gamma_\mu\gamma_5\frac{\tau_i}{2}\psi$, where $\tau_i$ is the $i$th Pauli sigma matrix acting on flavor space. With this definition, the physical pion decay constant is approximately $92~\mathrm{MeV}$. This definition implies
\ba \label{eq:Fpi-definition}
\bra{0}A^i_0(x,t) \ket{\pi(\mathbf{p}=0)}=F_\pi m_\pi e^{-m_\pi t}
\ea 
and
\ba \bra0 \partial_\mu A_\mu^i(x)\ket{\pi^i(\mathbf{p}=0)}=-m_\pi^2 F_\pi e^{-m_\pi t}.\ea 
We can calculate $F_\pi$ directly by fitting correlation functions of the axial current with pion fields. We can also use Noether's theorem to replace $\partial_\mu A_\mu^i(x)$ so that we can fit correlation functions involving only pion fields.

On the lattice, we modify the definition of $F_\pi$ to account for periodic boundary conditions. We define the axial current as the conserved current associated with rotations involving $\phi_0$. Specifically, we choose the discretization
\ba 
A_i^\mu \equiv \phi_0(x-\mu)\phi_i(x)-\phi_0(x)\phi_i(x-\mu).
\ea 
With this definition, the appropriate discrete version of Noether's theorem is
\ba 
\sum_\mu [A_i^\mu(x+\mu)-A_i^\mu(x)]=-\alpha \phi_i.
\ea 

\subsection{Isospin Channels}\label{sec_isospin_channel}

In QCD, we may define operators that couple to various pion states. For example, we could define  
\ba  \pi^\pm = \frac{i}{\sqrt{2}}\bar\psi\gamma^5\left(\pm\frac{\tau^1}{\sqrt{2}}+i\frac{\tau^2}{\sqrt{2}}\right)\psi\quad\text{and}\quad  \pi^0 = i\bar\psi\gamma^5\frac{\tau^3}{\sqrt{2}}\psi.\ea 
In our effective theory, we can build corresponding operators
\ba  \pi^\pm = \frac{1}{\sqrt{2}}\left(\pm\phi_1+i\phi_2\right)\quad\text{and}\quad  \pi^0=\phi_3.\ea 
From these operators, we can create two-pion operators with specific angular momentum and isospin. In this paper, we consider only operators that are angularly symmetric. For our isospin-0 operator, we choose
\ba \mathcal{O}^{I=0}(p)=\frac{1}{\sqrt{3}N_p}\sum_{|\mathbf{p}|=p}\left[ \pi^+(\mathbf{p}) \pi^-(-\mathbf{p})- \pi^0(\mathbf{p}) \pi^0(-\mathbf{p})+ \pi^-(\mathbf{p}) \pi^+(-\mathbf{p})\right]\ea 
\ba =-\frac{1}{\sqrt{3}N_p}\sum_{|\mathbf{p}|=p}\left[\phi_1(\mathbf{p})\phi_1(-\mathbf{p})+\phi_2(\mathbf{p})\phi_2(-\mathbf{p})+\phi_3(\mathbf{p})\phi_3(-\mathbf{p})\right].\ea 
The sum is over all momenta $\mathbf{p}$ on the lattice with norm $p$, and $N_p$ is the number of distinct lattice momenta that satisfy this condition. The pions that this operator couples to will have a relative momentum of $2p$. For our isospin-2 operator, we choose
\ba \mathcal{O}^{I=2}(p)=\frac{1}{\sqrt{6}N_{p}}\sum_{|\mathbf{p}|=p}\left[ \pi^+(\mathbf{p}) \pi^-(-\mathbf{p})+2 \pi^0(\mathbf{p}) \pi^0(-\mathbf{p})+ \pi^-(\mathbf{p}) \pi^+(-\mathbf{p})\right]\ea 
\ba =\frac{1}{\sqrt{6}N_{p}}\sum_{|\mathbf{p}|=p}\left[-\phi_1(\mathbf{p})\phi_1(-\mathbf{p})-\phi_2(\mathbf{p})\phi_2(-\mathbf{p})+2\phi_3(\mathbf{p})\phi_3(-\mathbf{p})\right].\ea

\section{Results} \label{results} 
\hl{All} the code to generate these results is available as part of the Qlattice library \cite{Qlattice}. \hl{A sketch of the overall workflow is as follows: 
\begin{itemize}
    \item We use the Hybrid Monte Carlo algorithm~\cite{Duane:1987de} to simulate the linear sigma model for various choices of the parameters $m^2$, $\lambda$, and $\alpha$ on a finite lattice with periodic boundary conditions. This results in an ensemble of random configurations of the scalar fields $\phi_i(x)$, with each random field configuration sampled according to the probability density $e^{-S_E[\phi]}$.
    \item We measure correlation functions of the scalar fields by approximating the path integral as a sum over our ensemble of field configurations
    \ba
    \langle T\{ \mathcal O[\phi_i] \} \rangle
    =
    \frac{1}{Z}\int D \phi_i\, \mathcal O[\phi_i] e^{-S_E[\phi]}
    \approx
    \frac{1}{N_\text{conf}}\sum_{\text{conf}}
    \mathcal O[\phi^{\text{conf}}_i],
    \ea
    where $N_\text{conf}$ is the number of configurations in the ensemble and we sum over the operator $\mathcal{O}$ evaluated on each configuration.
    \item We use the identifications of \Cref{sec_particle_ids,sec_pion_decay_constant} to connect these correlation functions to QCD matrix elements. See \Cref{sec:correlation_functions} for a discussion of how each observable was determined from the scalar field correlation functions.
    \item Using the same method used in Ref. \cite{RBC_pipi_scattering} for lattice QCD, we solve the generalized eigenvalue problem (GEVP) to get the energy spectrum of our theory in the isospin-0 and isospin-2 channels.
\end{itemize}
}

\subsection{Lambda-dependence}
\begin{figure}
	\centering
	\includegraphics[width=\textwidth]{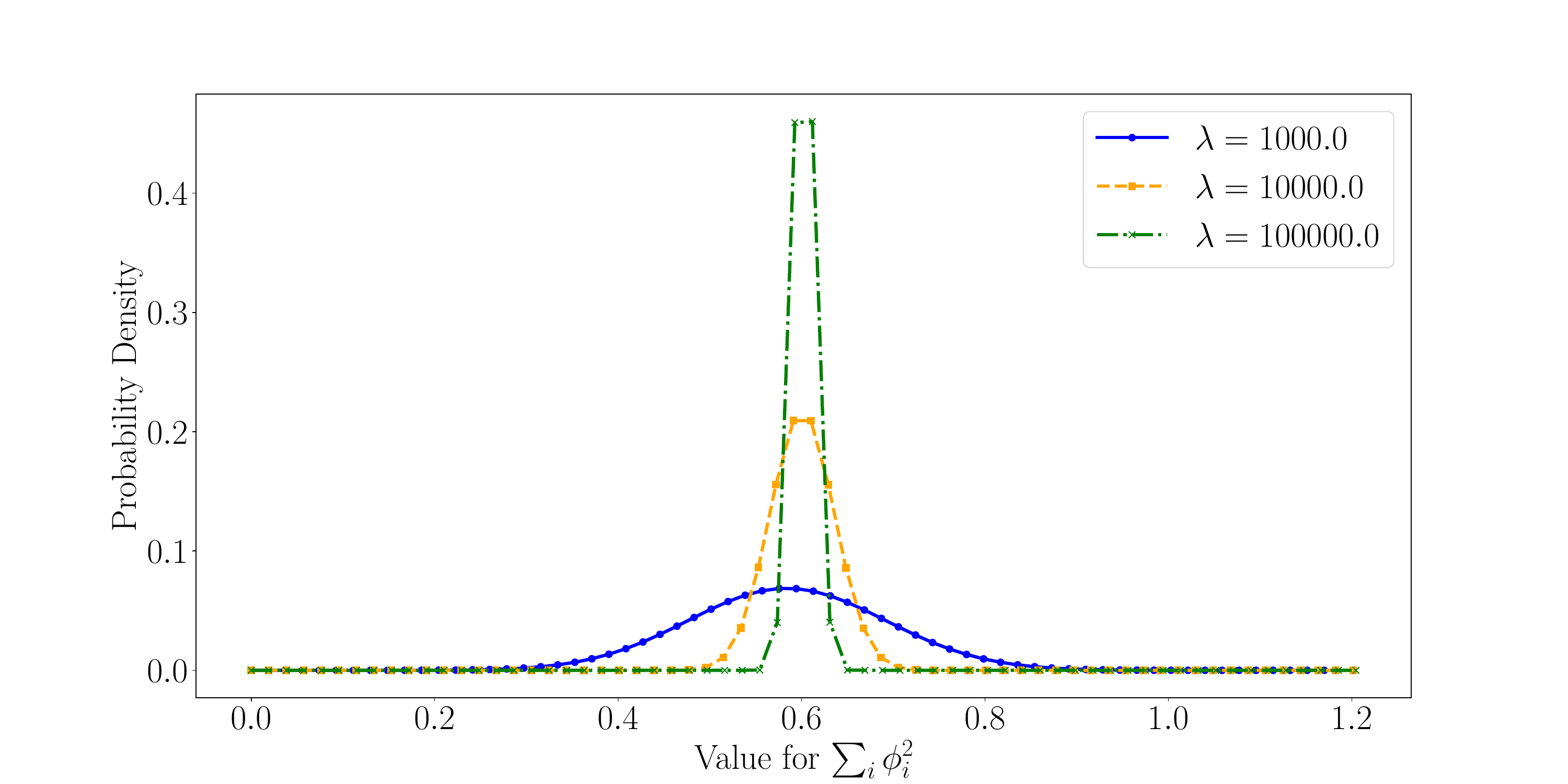}
	\caption{The probability distribution of $\sum_i\phi_i^2(x)$ for various $\lambda$ on a $16^3\times 32$ lattice with $m^2/\lambda=-0.102$ and $\alpha=0.007$. These parameters are chosen so that we are near the physical point (defined as the point where $m_\pi/F_\pi=135/92\approx 1.47$) when $\lambda=10^4$ (see \Cref{ensemble_parameters} for more information on the ensemble with $\lambda=10^4$). As can be seen from the figure, at large $\lambda$, the values of the fields are dynamically constrained to lie near the surface of a 3-sphere.}
	\label{phi_sq_dist}
\end{figure}

We want to study the linear sigma model in the large $\lambda$ limit, where it behaves like the non-linear sigma model. In \Cref{phi_sq_dist}, we plot the probability distribution of $\sum_{i=0}^3\phi_i^2(x)$ (by translational invariance, the point in space does not matter) for three different values of $\lambda$. In the $\lambda\to\infty$ limit, this distribution would be a delta function. These plots show that we have indeed chosen $\lambda$ sufficiently large that the fields are being dynamically constrained to lie close to the surface of a 3-sphere. We can therefore expect our theory to behave similarly to the non-linear sigma model.

\begin{figure}
	\centering
	\includegraphics[width=\textwidth]{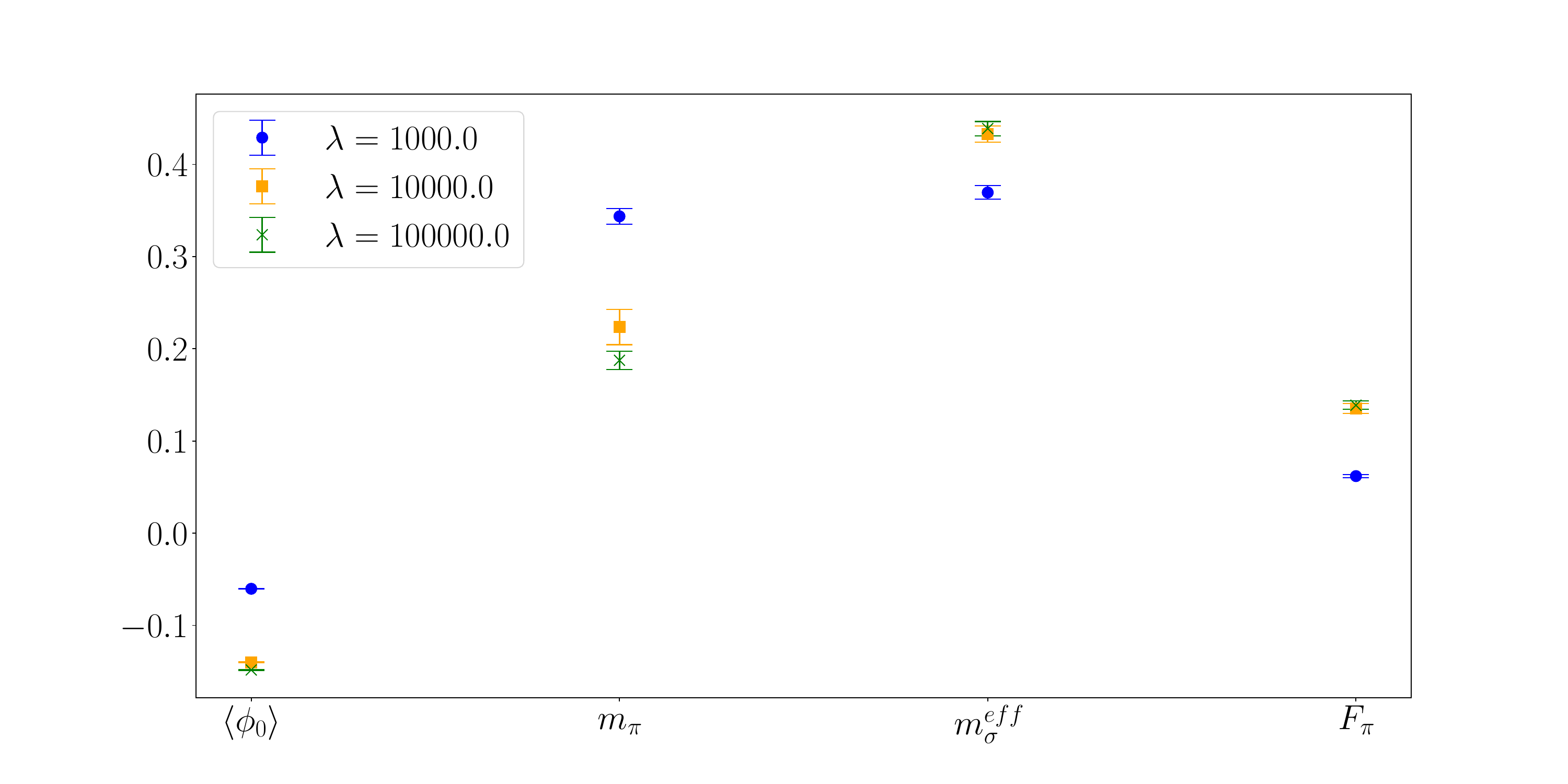}
	\caption{The $\lambda$-dependence of the vacuum expectation value of $\phi_0$, the pion mass $m_\pi$, the effective sigma mass $m_\sigma^\text{eff}$, and the pion decay constant $F_\pi$ on a $16^3\times 32$ lattice. All values are in lattice units. We set $m^2/\lambda=-0.102$ and $\alpha=0.007$ so that we are near the physical point (defined as the point where $m_\pi/F_\pi=135/92\approx 1.47$) when $\lambda=10^4$  (see \Cref{ensemble_parameters} for more information on the ensemble with $\lambda=10^4$). The effective sigma mass is based on the values of the $\sigma$-$\sigma$ correlation function at time-separations 0 and 1. As can be seen from the Figure, increasing $\lambda$ from $10^4$ to $10^5$ does not significantly affect $m_\sigma^\text{eff}$.}
	\label{lambda_comp}
\end{figure}

In \Cref{lambda_comp}, we show how varying $\lambda$ affects various observables. $\frac{m^2}{\lambda}$ and $\alpha$ are fixed so that we are near the physical ratio of $m_\pi$ to $F_\pi$ at $\lambda\approx 10^4$ (we define the physical ratio to be $m_\pi/F_\pi=135/92\approx 1.47$). Based on \Cref{lambda_comp}, we note that further increasing past $\lambda=10^4$ does not seem to have a significant effect on our results. Therefore, in subsequent sections, we will do all our calculations using $\lambda=10^4$. As $\lambda$ becomes larger, simulations become more difficult, since the dynamically-imposed constraint on $\sum_i\phi_i^2(x)$ becomes more stringent.

\subsection{The Phase Transition and Sigma Stability}
We fix $\lambda=10^4$. The two remaining parameters $m^2$ and $\alpha$ can be tuned to find the physical point. For our purposes, we define the physical point as the point where $m_\pi=135$ MeV and $F_\pi=92$ MeV. With this definition, we tune $m^2$ and $\alpha$ until the ratio $m_\pi/F_\pi$ is close to its physical value of $135/92\approx 1.47$. We then use the requirement $F_\pi=92$ MeV to define the lattice spacing.

\begin{figure}
	\centering
	\includegraphics[width=\textwidth]{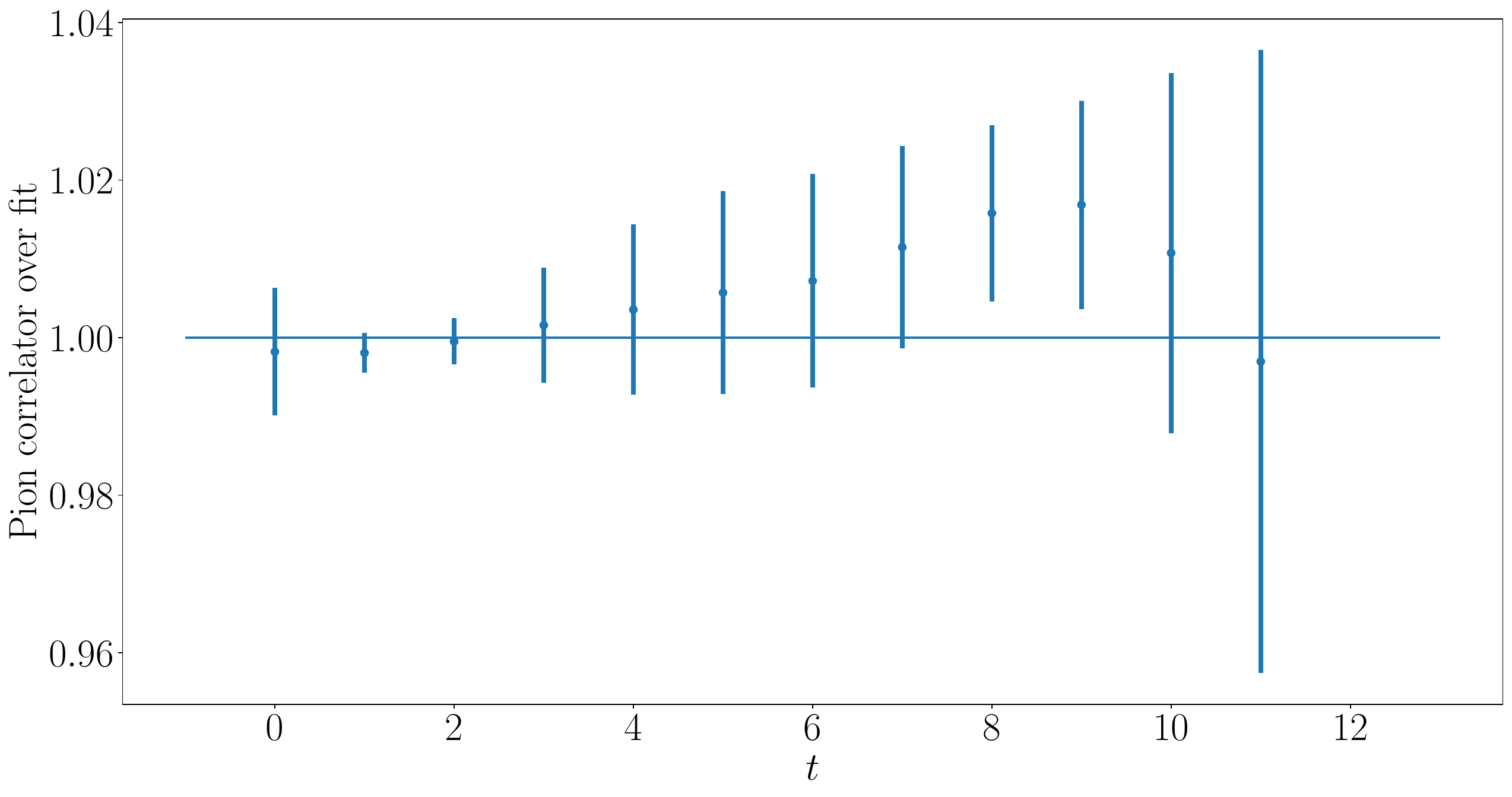}
	\caption{The correlation function $\left\langle \left(\sum_{i=1}^3\sum_\mathbf{x}\phi_i(\mathbf{x},t)\right)\left(\sum_{j=1}^3\sum_{\mathbf{x}'}\phi_j(\mathbf{x}',0)\right)\right\rangle$ divided by a single-state fit $f(t) = \frac{A}{\cosh((N_t/2-1)m_\pi)}\cosh\left(\left(\frac{N_t}{2}-t\right)m_\pi\right)$, where $A$ and $m_\pi$ are the fit parameters, and $N_t$ is the time extent of the lattice \hl{(see \Cref{sec:correlation_functions})}. This correlation function was calculated on the $16^3\times 128$ ensemble with $m^2/\lambda=-0.102$, $\alpha=0.007$, and $\lambda=10^4$ (see \Cref{gevp_ensembles} for more details on this ensemble). The fit gives $A=7238(102)$ and $m_\pi=0.2201(77)$. As can be seen from the figure, single-state fit works well even at small $t$.}
	\label{pipi_over_fit}
\end{figure}

When we calculate the pion mass, we fit correlation functions \hl{$\langle \mathcal{O}_\pi(t)\mathcal{O}_\pi(0)\rangle$ of the pion interpolating operator $\mathcal{O}_\pi(t)\equiv \sum_{i=1}^3\sum_\mathbf{x}\phi_i(\mathbf{x},t)$ (see \Cref{sec:correlation_functions})}. As can be seen from \Cref{pipi_over_fit}, a simple single-state fit works even for small $t$. \hl{This stands in contrast to lattice QCD, where excited states of the pion can couple to the pion interpolating operators (see for example \cite{Yan:2025mdm}). In our effective theory, the pions have no internal degrees of freedom. Since G-parity forbids $\mathcal{O}_\pi$ from coupling to two-pion states, the first excited state contributions would have to come from three-pion states, which are numerically very small.}

\begin{figure}
	\centering
	\includegraphics[width=0.95\textwidth]{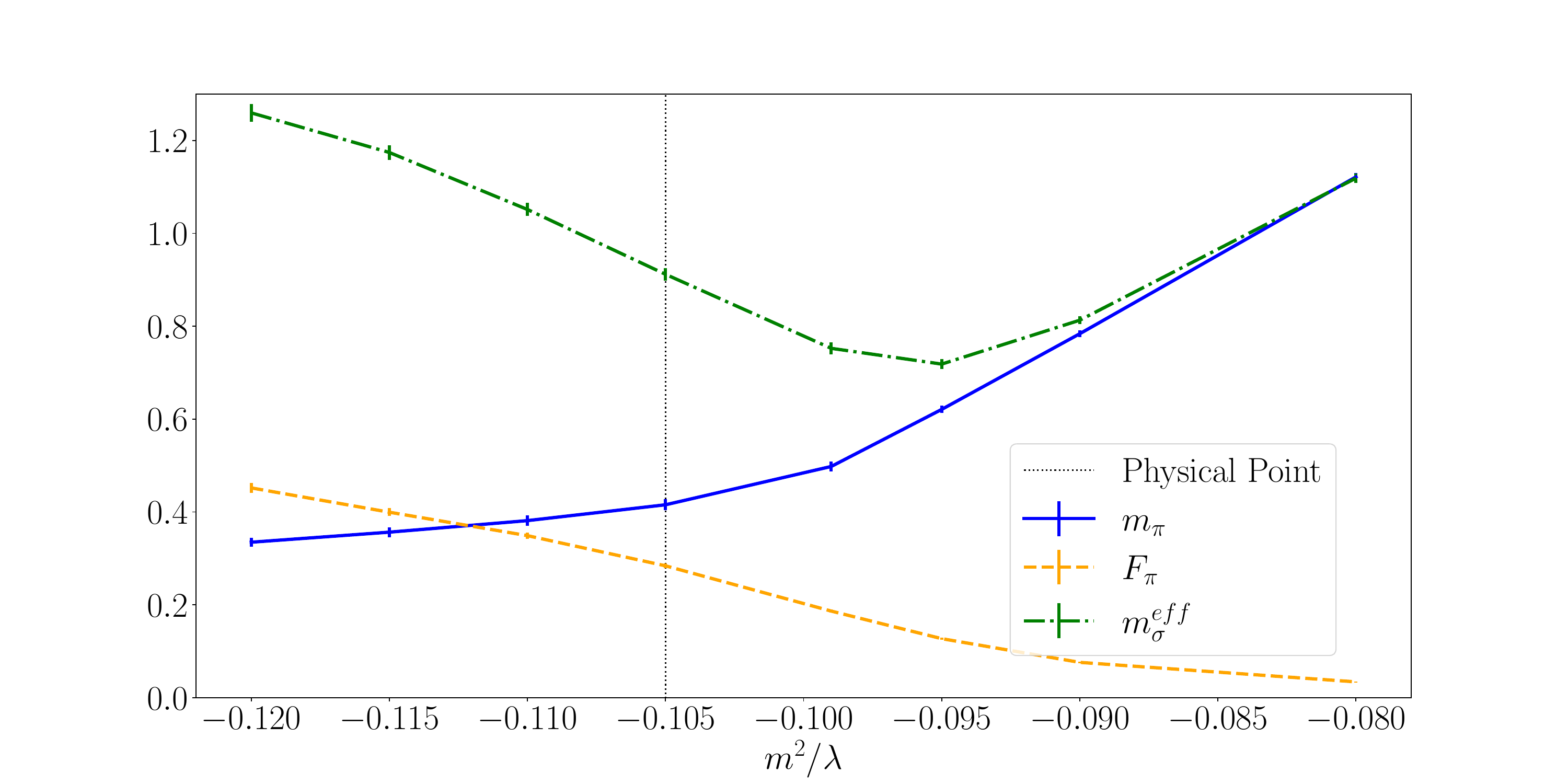}
	\includegraphics[width=0.95\textwidth]{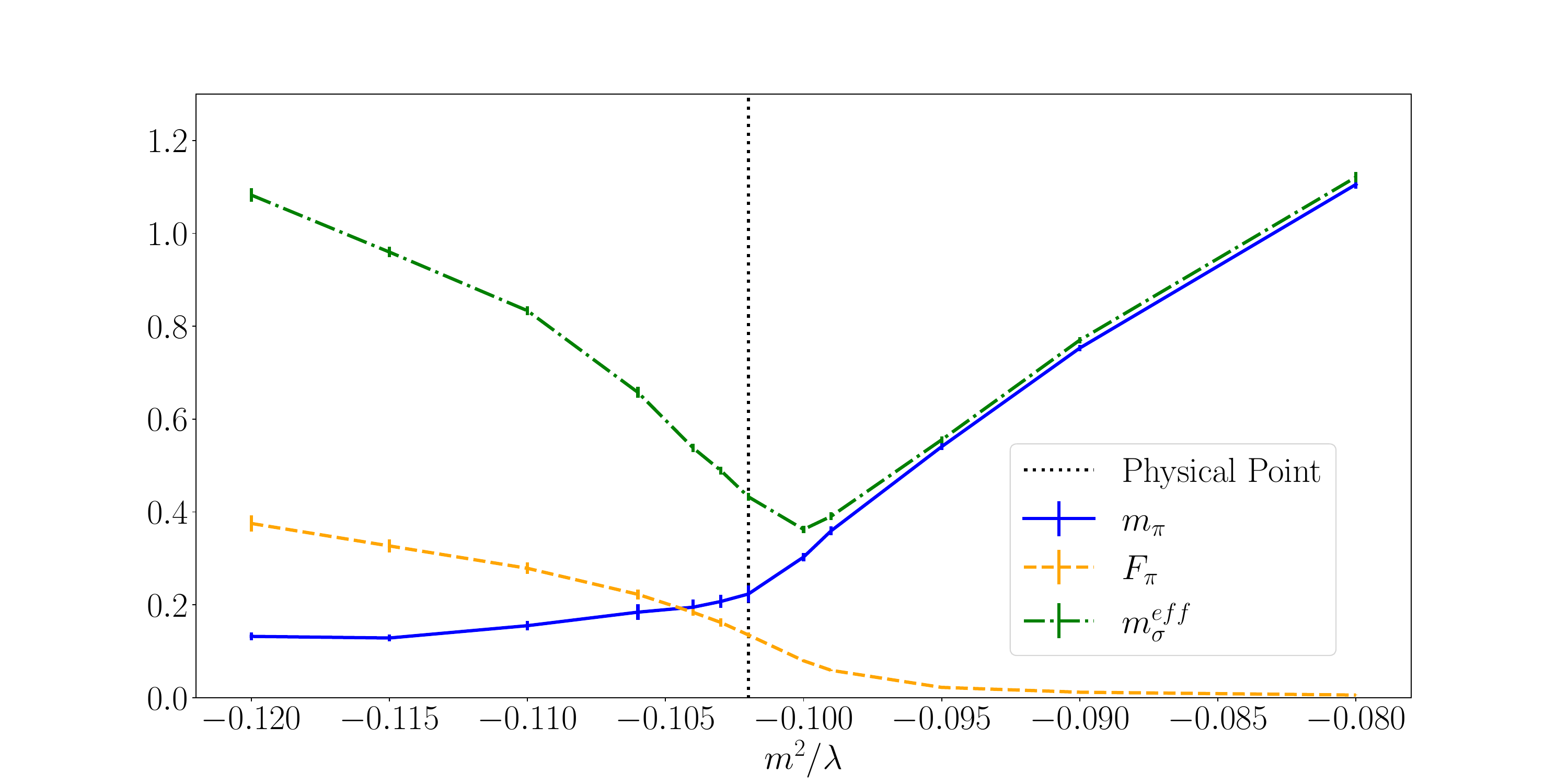}
	\caption{The pion mass $m_\pi$, the effective sigma mass $m_\sigma^\text{eff}$, and the pion decay constant $F_\pi$ (all measured in lattice units) versus $\frac{m^2}{\lambda}$ on a $8^3\times 16$ lattice (top) and a $16^3\times 32$ lattice (bottom). The value of $m^2/\lambda$ closest to the physical point (defined as the point where $m_\pi/F_\pi=135/92\approx 1.47$) is marked by a vertical line. $\lambda$ is held fixed at $10^4$, which we have determined is sufficiently large that further increases will not have a strong effect on the results. We set $\alpha=0.05$ for the $8^3\times 16$ lattice and $\alpha=0.007$ for the $16^3\times 32$ lattice. The resulting lattice spacing and spatial extent near the physical points are given in \Cref{ensemble_parameters}. The effective sigma mass is based on the values of the $\sigma$-$\sigma$ correlation function at time-separations 0 and 1. Increasing the magnitude of $\frac{m^2}{\lambda}$ causes $F_\pi$ to increase while $m_\pi$ decreases, making it clear that the physical point is unique.}
	\label{params_vs_msq}
\end{figure}

As shown in \Cref{params_vs_msq}, when the parameter ${m^2}/{\lambda}$ is sufficiently small in magnitude (in our convention, $m^2$ is negative), the theory is in the symmetric phase where $m_\pi\approx m_\sigma$ and $F_\pi \approx 0$. The theory enters the symmetry-broken phase around ${m^2}/{\lambda}\approx -0.1$, which is consistent with the results of previous lattice studies using the non-linear sigma model \cite{Hasenfratz_Higgs_bound_1988}. In the symmetry-broken phase, $F_\pi$ is no longer zero. Based on the tree-level formula $F_\pi=\sqrt{{-6m^2}/{\lambda}}$, we expect $F_\pi$ to continue increasing as we increase $\left|{m^2}/{\lambda}\right|$. This expectation is \hl{qualitatively} supported by our results in \Cref{params_vs_msq}\hl{, although we see significant departures from the tree-level formula}. On the other hand, $m_\pi$ starts with a large value in the symmetric phase ($m_\pi\approx m_\sigma$). As we transition into the symmetry-broken phase by increasing $\left|{m^2}/{\lambda}\right|$, $m_\pi$ decreases significantly because it is the mass of a pseudo-Goldstone boson. Based on \Cref{params_vs_msq}, $m_\pi$ continues to decrease after the phase transition as we increase $\left|{m^2}/{\lambda}\right|$.

Since we have $F_\pi$ increasing and $m_\pi$ decreasing as a function of $\left|{m^2}/{\lambda}\right|$, there is only one physical point for each given value of $\alpha$. \hl{Note that we cannot tune $F_\pi$ and $m_\pi$ independently of each other, since a change in either $m^2/\lambda$ or $\alpha$ affects both $F_\pi$ and $m_\pi$, as shown in \Cref{params_vs_msq}}. In fact, in all our calculations, the physical point occurs close to the symmetric phase at roughly the same value of ${m^2}/{\lambda}$, as shown in \Cref{ensemble_parameters}. The parameter $\alpha$ serves mainly to control the lattice spacing.

As can be seen from \Cref{params_vs_msq}, at the physical point, the effective sigma mass is about twice the pion mass, indicating that the sigma particle is on the border between being stable and unstable. Furthermore, if we were to simulate larger-than-physical pion masses, holding the pion decay constant at the same value, then the sigma would become a stable particle.

\begin{table}
	\centering
	\begin{tabular}{|| c | c | c | c | c | c | c | c ||} 
		\hline
        Size & $m^2/\lambda$ & $\alpha$ & $a^{-1}$ (GeV) & $L$ (fm) & $m_\pi a$ & $F_\pi a$ & $m_\pi/F_\pi$ \\
		\hline\hline
		$8^3\times 16$ & -0.105 & 0.05 & $0.3237(61)$ & 4.876(92) & 0.416(12) & 0.2841(54) & 1.462(43) \\ 
		\hline
		$12^3\times 24$ & -0.103 & 0.015 & $0.474(12)$ & 4.99(13) & 0.272(11) & 0.1940(50) & 1.427(59) \\ 
		\hline
		$16^3\times 32$ & -0.102 & 0.007 & $0.679(28)$ & 4.64(19) & 0.223(19) & 0.1353(56) & 1.65(14) \\ 
		\hline
	\end{tabular}
	\caption{Parameters and observables \sout{on various}\hl{for the } ensembles \hl{from \Cref{params_vs_msq}} \sout{near}\hl{closest to} the physical point (defined as the point where $m_\pi/F_\pi=135/92\approx 1.47$). The inverse lattice spacing $a^{-1}$ is defined by requiring $F_\pi=92$ MeV. $L$ is \hl{the length of the lattice volume along each of the three spatial dimensions}. As usual, $\lambda = 10^4$ in all these ensembles.}
	\label{ensemble_parameters}
\end{table}

\begin{table}
	\centering
	\begin{tabular}{|| c | c | c | c | c | c | c | c ||} 
		\hline
        Size & $m^2/\lambda$ & $\alpha$ & $a^{-1}$ (GeV) & $L$ (fm) & $m_\pi a$ & $F_\pi a$ & $m_\pi/F_\pi$ \\
		\hline\hline
		$16^3\times 128$ & -0.102 & 0.007 & 0.6363(99) & 4.961(77) & 0.2201(77) & 0.1445(23) & 1.522(59) \\ 
		\hline
		$8^3\times 128$ & -0.105 & 0.05 & 0.3323(18) & 4.750(26) & 0.4196(43) & 0.2769(15) & 1.516(19) \\ 
		\hline
		$12^3\times 128$ & -0.105 & 0.05 & 0.3271(19) & 7.238(41) & 0.4151(35) & 0.2812(16) & 1.476(15) \\ 
		\hline
		$16^3\times 128$ & -0.105 & 0.05 & 0.3272(26) & 9.649(77) & 0.4118(74) & 0.2812(22) & 1.464(30) \\ 
		\hline
	\end{tabular}
	\caption{Parameters and observables for the ensembles used for the GEVP analysis. The inverse lattice spacing $a^{-1}$ is defined by requiring $F_\pi=92$ MeV. $L$ is \sout{the spatial extent of the lattice}\hl{the length of the lattice volume along each of the three spatial dimensions}. The physical value of $m_\pi/F_\pi$ is about $135/92\approx 1.47$. As usual, $\lambda = 10^4$ in all these ensembles.
    }
	\label{gevp_ensembles}
\end{table}

\subsection{Energy Spectrum from GEVP}

\hl{
We can determine the finite-volume spectrum of our theory by solving a generalized eigenvalue problem (GEVP). We create a list of operators $\{\mathcal{O}_i\}$ with a particular set of quantum numbers (e.g. $I=0$) and define corresponding operator states
\ba 
|\mathcal{O}_i\rangle\equiv \frac{1}{N_i}(\mathcal{O}_i-\langle\mathcal{O}_i\rangle)|0\rangle,
\ea 
where $\langle\mathcal{O}_i\rangle$ is the vacuum expectation value of $\mathcal{O}_i$ and $N_i$ is chosen to normalize the state to 1. We then approximate the energy eigenstates by writing them as linear combinations of the operator states
\ba \label{eq:GEVP_eigenstates}
| E_{I,n} \ra
=
\sum_i
c_{I,n}^{i} | \mathcal O_i \ra.
\ea
We can determine the energy eigenvalues $e^{-E_{I,n}}$ and the coefficients $c^i_{I,n}$ by requiring that
\ba \label{eq:GEVP}
\langle E_{I,n} |
e^{-H(t+1)}
| \mathcal O_j \ra
=&
e^{-E_{I,n}}
\langle E_{I,n} |
e^{-Ht}
| \mathcal O_j \ra
\ea
for all $j$.
}

\hl{
For the $I=0$ GEVP, we have chosen the set of operators 
\ba
&\mathcal{O}_\pi^{I=0}(\mathbf{p}=0) = -\frac{1}{\sqrt{3}}\sum_{i=1}^3[\phi_i(\mathbf{p}=0)]^2
\\
&\mathcal{O} = -\phi_0(\mathbf{p}=0)
\\
&\mathcal{O}_\pi^{I=0}\left(\mathbf{p}=\frac{2\pi}{L}\right) = -\frac{1}{6\sqrt{3}}\sum_{|\mathbf{p}|={2\pi}/{L}}\sum_{i=1}^3\left[\phi_i(\mathbf{p})\phi_i(-\mathbf{p})\right],
\ea
where $\phi_i(\mathbf{p})\equiv \sum_{\mathbf{x}} \phi_i(\mathbf{x})e^{-i\mathbf{p}\cdot\mathbf{x}}$. For the $I=2$ GEVP, we have chosen 
\ba 
&\mathcal O_\pi^{I=2}(\mathbf{p}=0) = \frac{1}{\sqrt{6}}\left(-[\phi_1(\mathbf{p=0})]^2-[\phi_2(\mathbf{p}=0)]^2+2[\phi_3(\mathbf{p}=0)]^2\right)
\\
&\mathcal O_\pi^{I=2}\left(\mathbf{p}=\frac{2\pi}{L}\right) = \frac{1}{6\sqrt{6}}\sum_{|\mathbf{p}|={2\pi}/{L}}\left[-\phi_1(\mathbf{p})\phi_1(-\mathbf{p})-\phi_2(\mathbf{p})\phi_2(-\mathbf{p})+2\phi_3(\mathbf{p})\phi_3(-\mathbf{p})\right].
\ea
See \Cref{sec_isospin_channel} for the explanation behind the assignment of isospin to these operators.
}

\begin{figure}
	\centering
	\includegraphics[width=0.49\textwidth]{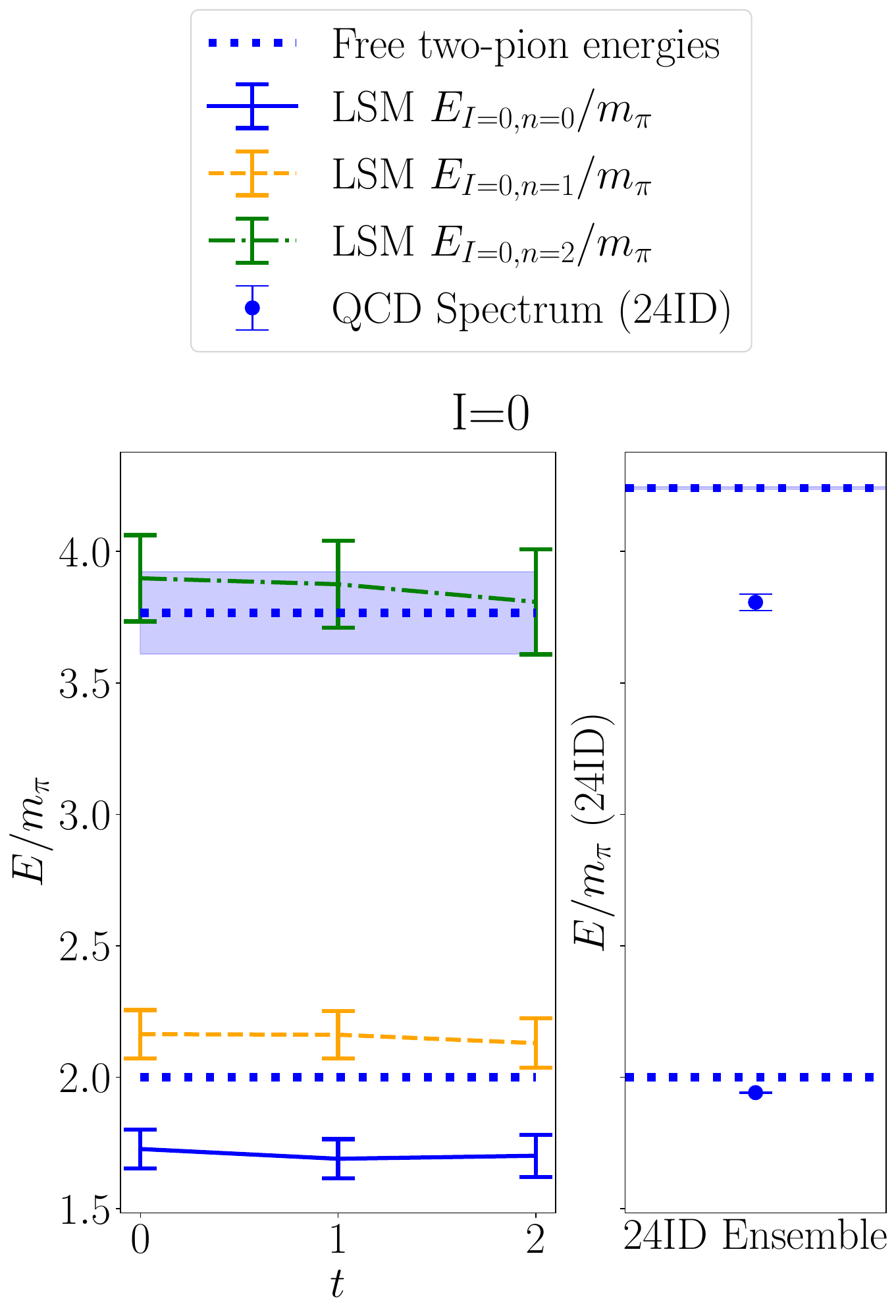}
	\includegraphics[width=0.49\textwidth]{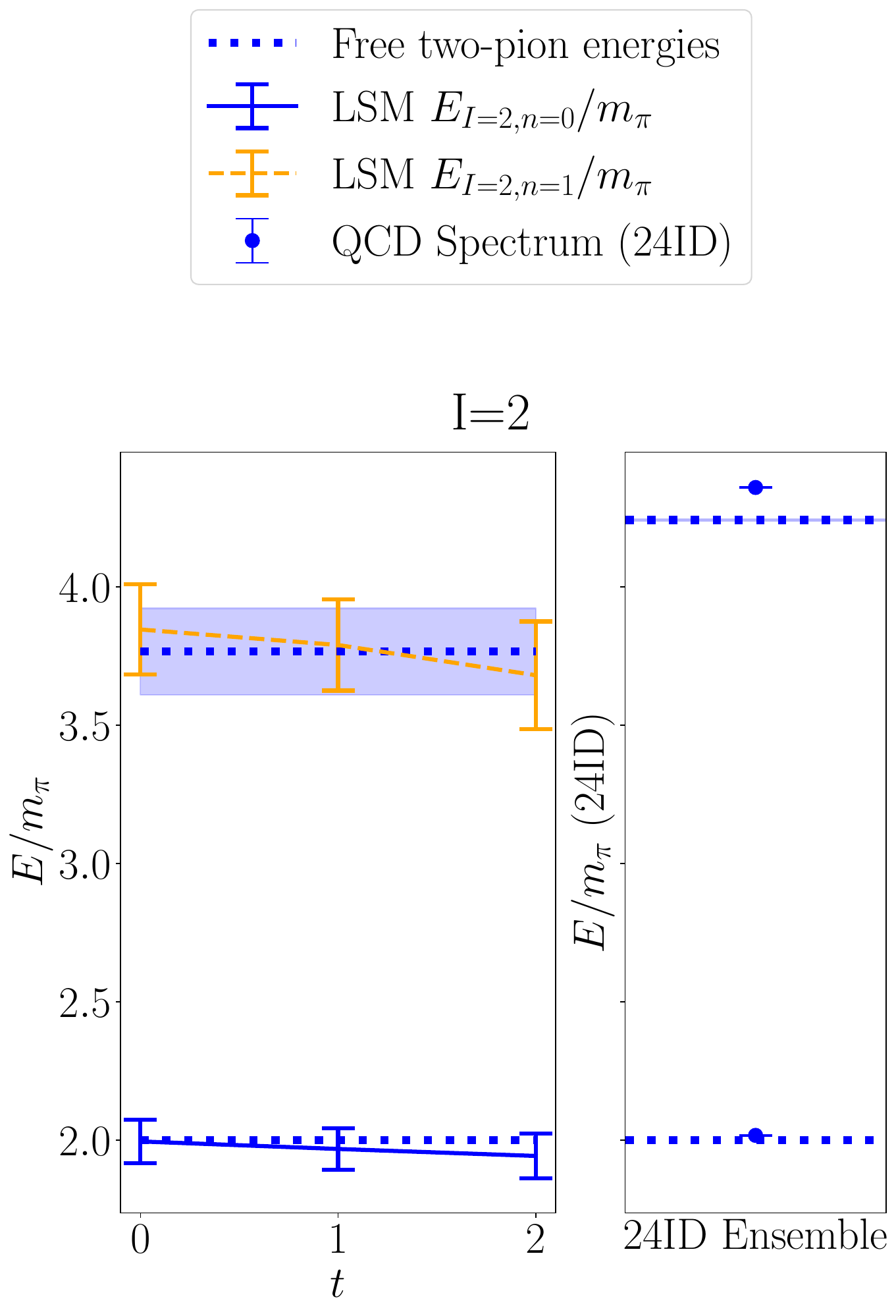}
	\caption{The isospin-0 (left) and isospin-2 (right) energy spectrum\hl{. The dotted lines show the energies of a system of two free (non-interacting) pions with either zero momentum or one unit of momentum. The orange, green, and solid blue lines give the energy eigenvalues for the linear sigma model} as determined by solving the \sout{generalized eiegenvalue problem}\hl{GEVP from \Cref{eq:GEVP}} using \sout{timeslices}\hl{time separations} $t$ and $t+1$ on a $16^3\times 128$ lattice with $m^2/\lambda = -0.102$ and $\alpha=0.007$ (as usual, $\lambda=10^4$). These parameters are chosen so that we are near the physical point with $a^{-1}=0.6363(99)$ GeV and $L=4.961(77)\text{ fm}$ (see \Cref{gevp_ensembles} for more details). \hl{The data points with error bars and marked with circles show} the QCD spectrum, as calculated \hl{on the 24ID ensemble} in \cite{RBC_pipi_scattering} on a $24^3\times 64$ lattice with similar physical volume ($L=4.63(1)$ fm) and $a^{-1}=1.023(2)$ GeV. The error bars on the QCD spectrum do not include the error in determining $m_\pi$.
    }
	\label{gevp_I0_I2}
\end{figure}

In \Cref{gevp_I0_I2}, we show the energy spectrum of the theory in the isospin-0 and isospin-2 channels and compare it to the spectrum in Ref.~\cite{RBC_pipi_scattering} from lattice QCD. To get our results, we solved the \sout{generalized eigenvalue problem at various time separations}\hl{GEVP from \Cref{eq:GEVP} for various choices of time separations $t$}. \sout{For the isospin-0 channel, we use a three-operator basis involving two-pion operators $\mathcal{O}^{0,0}(p=0)$ and $\mathcal{O}^{0,0}\left(p=\frac{2\pi}{L}\right)$, with relative momenta zero and $\frac{4\pi}{L}$ respectively, and the sigma operator $\phi_0$. For the isospin-2 channel, we use a two-operator basis $\mathcal{O}^{2,0}(p=0)$ and $\mathcal{O}^{2,0}\left(p=\frac{2\pi}{L}\right)$.}We chose the lattice spacing so that the volume of the lattice \hl{(as measured in physical units)} is about the same as in Ref.~\cite{RBC_pipi_scattering} (see \Cref{spectrum_comp} for a comparison).

\begin{figure}
	\centering
	\includegraphics[width=0.49\textwidth]{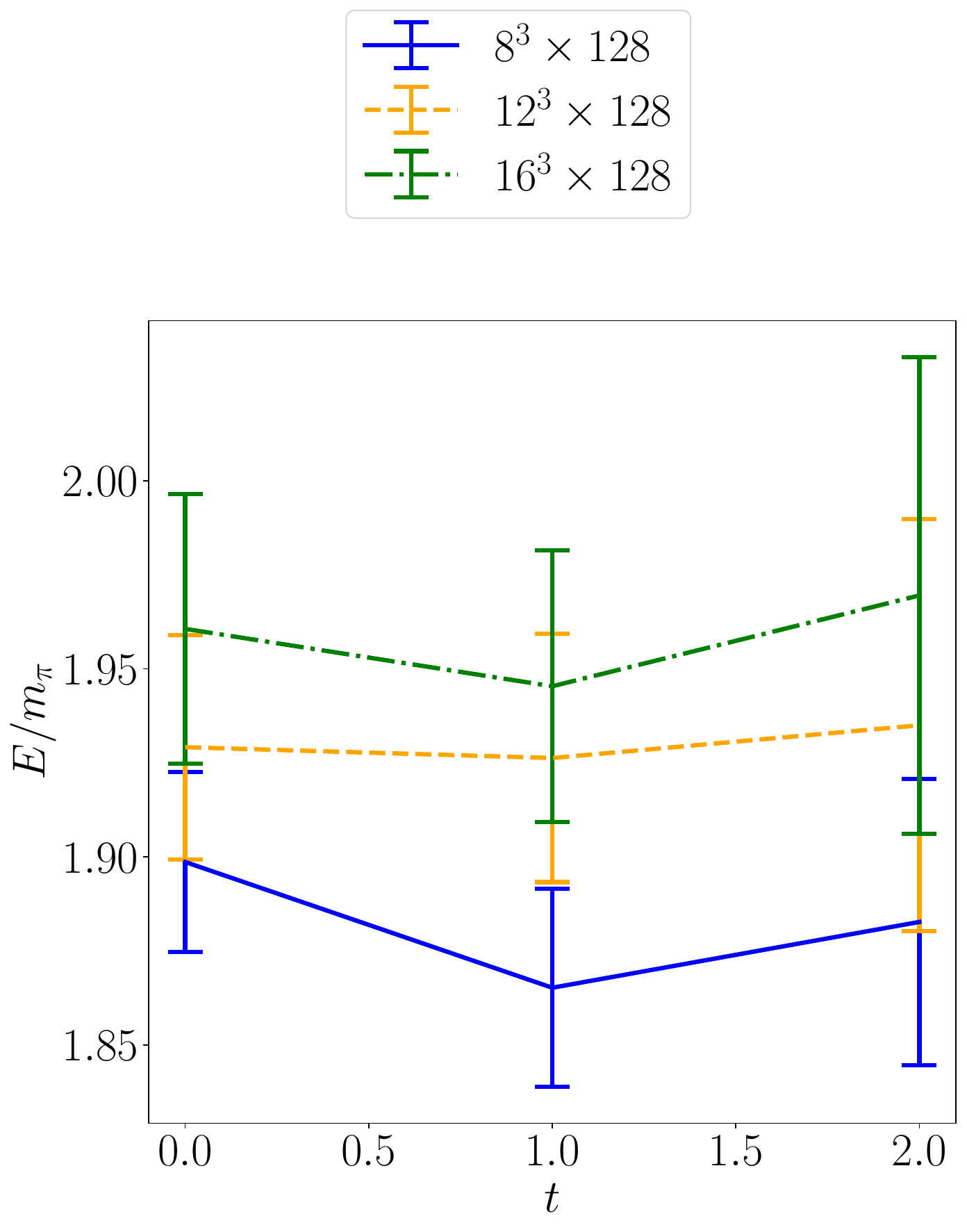}
	\includegraphics[width=0.49\textwidth]{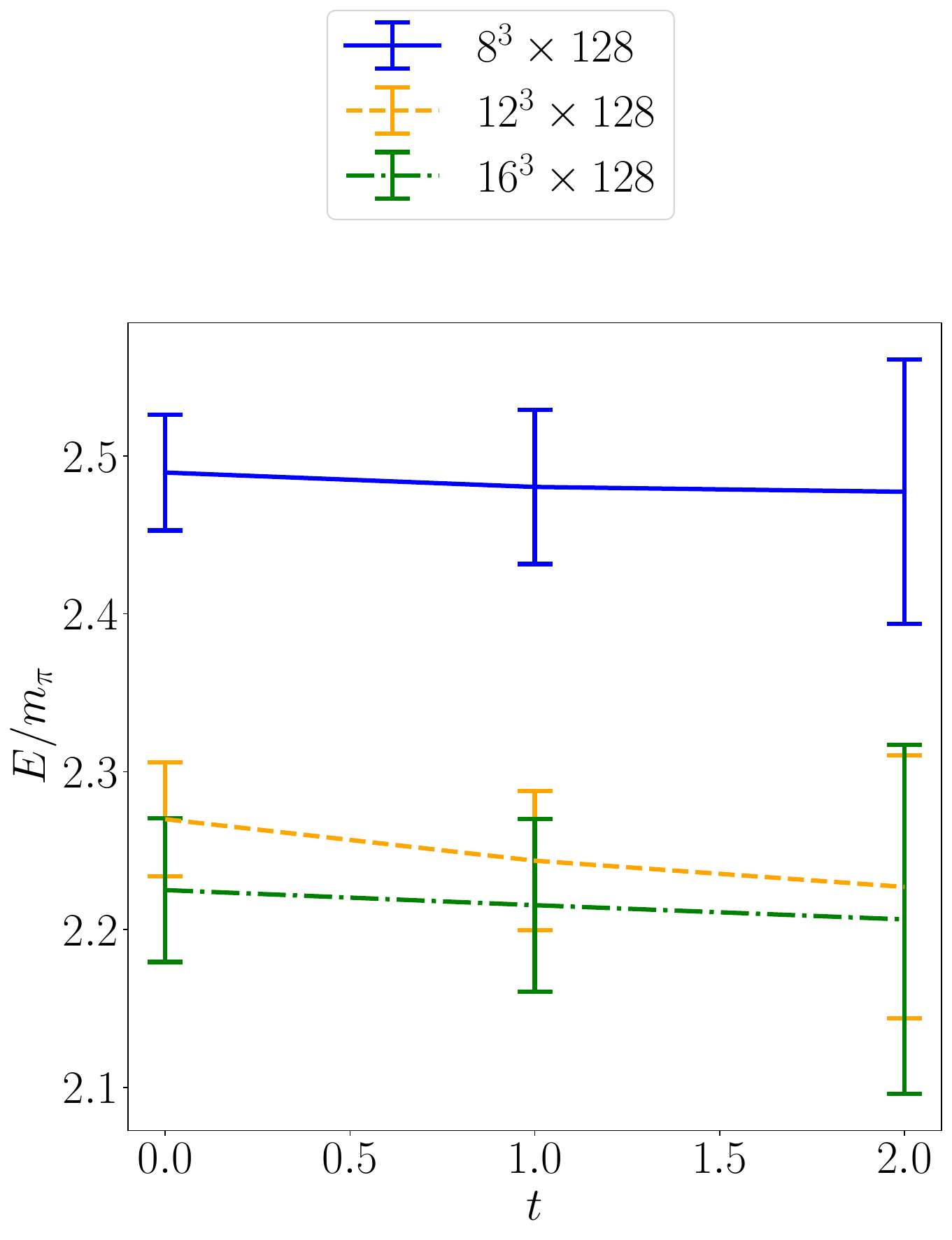}
	\caption{The isospin-0 ground state (left) and first excited state (right), as determined by solving the generalized eigenvalue problem using \sout{timeslices}\hl{time separations} $t$ and $t+1$, for three different lattice volumes. The simulations are done with $m^2/\lambda=-0.105$ and $\alpha=0.05$ (as usual, $\lambda=10^4$) so that we are near the physical point (see \Cref{gevp_ensembles} for more details on these ensembles).}
	\label{gevp_vol}
\end{figure}

In \Cref{gevp_vol}, we show the volume-dependence of the \hl{$I=0$} ground state energy and first excited state energy respectively. Without changing any of the parameters in the Lagrangian, we simulate at different spatial volumes, starting with our $8^3\times 128$ lattice with $L = 4.750(26)\text{ fm}$. As the volume increases, the ground state energy increases\hl{,} and\hl{, as shown in \Cref{eigenstate_0_volume_dependence}, the ground state} becomes more associated with the $\pi\pi$ operator at zero relative momentum\sout{, while}\hl{. On the other hand, as the volume increases,} the first excited state energy decreases\hl{,} and\hl{, as shown in \Cref{eigenstate_1_volume_dependence}, the first excited state} becomes more associated with the $\sigma$ operator. \hl{The second excited state is mostly associated with the $\pi\pi$ operator with one unit of momentum for each pion, as shown in \Cref{eigenstate_2_volume_dependence}.}

\begin{table}
	\centering
    \hl{
	\begin{tabular}{|| c | c | c | c ||} 
		\hline
        Lattice Volume & $c_{I=0,n=0}^{\pi,\mathbf{p}=0}$ & $c_{I=0,n=0}^\sigma$ & $c_{I=0,n=0}^{\pi,\mathbf{p}={2\pi}/{L}}$ \\
        \hline\hline
        $8^3\times 128$ & 0.8110(92) & -0.585(13) & 0.0219(53) \\
        \hline
        $12^3\times 128$ & 0.9137(81) & -0.407(18) & 0.0185(80) \\
        \hline
        $16^3\times 128$ & 0.9466(84) & -0.321(24) & 0.046(16)\\
		\hline
	\end{tabular}
	\caption{The coefficients $c_{I,n}^i$ from \Cref{eq:GEVP_eigenstates} for the $I=0$ ground state, as determined by the GEVP analysis using time separations 0 and 1. The simulations are done with $m^2/\lambda=-0.105$ and $\alpha=0.05$ (as usual, $\lambda=10^4$) so that we are near the physical point (see \Cref{gevp_ensembles} for more details on these ensembles).}
    \label{eigenstate_0_volume_dependence}
    }
\end{table}

\begin{table}
	\centering
    \hl{
	\begin{tabular}{|| c | c | c | c ||} 
		\hline
        Lattice Volume & $c_{I=0,n=1}^{\pi,\mathbf{p}=0}$ & $c_{I=0,n=1}^\sigma$ & $c_{I=0,n=1}^{\pi,\mathbf{p}={2\pi}/{L}}$ \\
        \hline\hline
        $8^3\times 128$ & 0.6269(89) & 0.7740(72) & -0.0902(81) \\
        \hline
        $12^3\times 128$ & 0.465(16) & 0.8708(89) & -0.163(13) \\
        \hline
        $16^3\times 128$ & 0.374(24) & 0.899(10) & -0.232(18) \\
		\hline
	\end{tabular}
	\caption{The coefficients $c_{I,n}^i$ from \Cref{eq:GEVP_eigenstates} for the $I=0$ first excited state, as determined by the GEVP analysis using time separations 0 and 1. The simulations are done with $m^2/\lambda=-0.105$ and $\alpha=0.05$ (as usual, $\lambda=10^4$) so that we are near the physical point (see \Cref{gevp_ensembles} for more details on these ensembles).}
	\label{eigenstate_1_volume_dependence}
    }
\end{table}

\begin{table}
	\centering
    \hl{
	\begin{tabular}{|| c | c | c | c ||} 
		\hline
        Lattice Volume & $c_{I=0,n=2}^{\pi,\mathbf{p}=0}$ & $c_{I=0,n=2}^\sigma$ & $c_{I=0,n=2}^{\pi,\mathbf{p}={2\pi}/{L}}$ \\
        \hline\hline
        $8^3\times 128$ & 0.1129(72) & 0.2341(76) & 0.9657(24) \\
        \hline
        $12^3\times 128$ & 0.0923(85) & 0.265(12) & 0.9600(36) \\
        \hline
        $16^3\times 128$ & 0.057(16) & 0.308(17) & 0.9499(57) \\
		\hline
	\end{tabular}
	\caption{The coefficients $c_{I,n}^i$ from \Cref{eq:GEVP_eigenstates} for the $I=0$ second excited state, as determined by the GEVP analysis using time separations 0 and 1. The simulations are done with $m^2/\lambda=-0.105$ and $\alpha=0.05$ (as usual, $\lambda=10^4$) so that we are near the physical point (see \Cref{gevp_ensembles} for more details on these ensembles).}
	\label{eigenstate_2_volume_dependence}
    }
\end{table}

\subsection{Comparison with Other Methods}
The spectrum for the isospin-2 channel is similar to the spectrum of full QCD, as recently calculated on the lattice by the RBC/UKQCD collaboration \cite{RBC_pipi_scattering}. In the isospin-0 channel, however, the spectrum looks very different. The results from our theory and lattice QCD are compared side-by-side in \Cref{spectrum_comp}. In our theory, there are two states at around twice the pion mass. These states are formed mainly from linear combinations of the $I=0$ two-pion operator $\mathcal{O}_\pi^{I=0}(p=0)$ and the sigma operator. Our second excited state is associated mainly with the operator $\mathcal{O}_\pi^{I=0}\left(p=\frac{2\pi}{L}\right)$. It occurs at around the same energy as the first excited state in lattice QCD. We interpret this spectrum as indicating the sigma resonance happens at a much lower energy in our theory than in QCD. \hl{Lattice QCD shows that the sigma becomes stable when the pion mass is above $300$ MeV \cite{Doring:2016bdr, Guo:2018zss, Rodas:2023gma}.}

We also note that our results show a different \hl{behavior for the sigma resonance than}\sout{dependence of the sigma resonance on the pion mass than is obtained from} unitarized chiral perturbation theory. Unitarized two-loop chiral perturbation theory predicts a sigma pole location of around 3 times the pion mass at the physical point, with the sigma becoming more stable as the pion mass increases until it finally becomes stable when the pion mass is around twice its physical value \cite{Pelaez_two-loop_2010}. In contrast, in our results, the sigma is already nearly stable at the physical point.

Finally, our results also differ from previously published results on the unitarized linear sigma model. Achasov and Shestakov \cite{Achasov_LSM_1994} calculated the pole mass and width of the sigma resonance by unitarizing the two-flavor linear sigma model, using parameters that correspond to $\lambda\approx 300$ in our conventions. They set their other parameters so that the tree-level values of \sout{$f_\pi$}\hl{$F_\pi$} and $m_\pi$ are physical. They then employ a simple unitarization model and show that it gives results for the $\pi\pi$ phase shift that are consistent with experiment at low energies. They calculate the sigma pole mass to be around $417$~MeV, in contrast to our results.

\begin{table}
	\centering
	\begin{tabular}{|| c || c | c ||} 
		\hline
		  & Linear sigma model & Lattice QCD \cite{RBC_pipi_scattering,Tu:2020vpn} \\
		\hline\hline
		$E_{I=0,n=0}/m_\pi$ & $1.727(73)$ & $1.9413(29)$* \\ 
		$E_{I=0,n=1}/m_\pi$ & $2.164(92)$ & $3.806(32)$* \\
		$E_{I=0,n=2}/m_\pi$ & $3.898(164)$ & $4.984(86)$* \\
		\hline
		$E_{I=2,n=0}/m_\pi$ & $1.995(78)$ & $2.0174(24)$* \\ 
		$E_{I=2,n=1}/m_\pi$ & $3.846(163)$ & $4.3595(22)$* \\
        \hline\hline
        $2E_{\pi,p=(2\pi/L)}/m_\pi$ & $3.767(156)$ & $4.242(4)$* \\
        \hline
        $m_\pi L$ & 3.52(12) & 3.3466(43) \\
        \hline
        $L/\mathrm{fm}$ & $4.961(77)$ & $4.63(1)$ \\
        \hline
        $a^{-1}/\mathrm{GeV}$ & $0.6363(99)$ & $1.023(2)$ \\
        \hline
        $m_\pi / \mathrm{MeV}$ & 140(5) & 142.6(3) \\
        \hline
        $F_\pi/\mathrm{MeV}$ & $92$ & $94.44(8)^\dagger$ \\
		\hline
	\end{tabular}
	\caption{The spectrum of the SU(2) linear sigma model versus QCD. $E_{I,n}$ is the energy of the $n$th excited state in the isospin-$I$ channel. \hl{$a$ is the lattice spacing, and $L$ is the length of the lattice volume along each of the three spatial dimensions.} The linear sigma model results are based on a GEVP analysis using \sout{timeslices}\hl{time separations} 0 and 1 on a $16^3\times 128$ lattice with $m^2/\lambda = -0.102$ and $\alpha=0.007$ (as usual, $\lambda=10^4$). These parameters are chosen so that we are near the physical point (defined as the point where $m_\pi/F_\pi=135/92\approx 1.47$). Non-interacting pion energy $2E_{\pi,p=2\pi/L}$ is shown for comparison. *These errors on the QCD spectrum do not include the error in determining $m_\pi$. ${}^\dagger$This error does not include the error in determining the lattice spacing.}
	\label{spectrum_comp}
\end{table}

\section{Power Counting with a Lattice Regularization}\label{power_counting}
In chiral perturbation theory (ChPT) with dimensional regularization, all corrections from loops and higher-order terms in the Lagrangian are suppressed by powers of $E/(4\pi F_0)$, where $E$ is the relevant energy scale and $F_0$ is the tree-level pion decay constant. This allows ChPT to remain predictive for energies below \sout{$4\pi f_\pi\approx 1$}\hl{$4\pi F_\pi\approx 1$} GeV. Now if we instead study ChPT with lattice regularization, we will have power divergences in addition to the logarithmic divergences that are present in dimensional regularization. As a result, we are no longer guaranteed that divergences will be suppressed by powers of $E/(4\pi F_0)$. For example, as calculated in Ref.~\cite{Shushpanov_lat_chpt_1998}, with lattice regularized ChPT, there is a quadratic divergence from the one-loop correction to the pion mass squared:
\begin{align}
    \frac{M^2 - M_0^2}{M_0^2} \approx \frac{0.2}{(F_0 a)^2} + \cdots
\end{align}
where $M_0$ is the tree-level pion mass. We see that the suppression of this divergence requires $(F_0 a)^2 \gg 0.2 $. To avoid large discretization effects, we must have $a\ll \frac{1}{M_0}$. Therefore, we must have $\frac{F_0}{M_0}\gg F_0 a\gg \sqrt{0.2}\approx 0.45$. Since the physical value of $\frac{F_\pi}{M_\pi}$ is about $92/135\approx 0.7$, we have good reason to be worried about the validity of power-counting arguments in lattice ChPT.

From our simulation results, we get non-perturbative confirmation that we are not justified in ignoring higher order terms in the effective theory expansion. Our leading order terms are not capable of reproducing the low-energy physics of QCD. We could of course try to add more terms to the Lagrangian. However, even if we could get the effective theory expansion to converge on the correct physics with more terms, we would at best be relying on delicate cancellations between the lower-order and higher-order terms.

Since the origin of this problem is the lattice regularization, a potential solution would be to change the regularization so that power divergences are more strongly suppressed. For example, we could smear the pion fields over adjacent lattice sites, making the lattice cutoff less harsh.

\section{Conclusion}
We simulated the O(4) linear sigma model on the lattice. These simulations were done at large values of the coupling constant, so that our theory behaves similarly to the non-linear sigma model. We demonstrated that our coupling constant was sufficiently large by noting that the fields were dynamically confined to lie close to the surface of a 3-sphere, and that further increases in the coupling constant had little effect on our observables. We found that the physical point is near the phase transition between the symmetric and symmetry-broken phases.

We calculated the spectrum of our theory in the isospin-0 and isospin-2 channels.
\hl{In the isospin-2 channel, we get similar results to those from lattice QCD. The first two states are two pions at rest and two pions with one unit of momentum each, with energies similar to what we would get for non-interacting pions. However,} we found that our results for the isospin-0 channel are very different from the same channel in QCD. In our calculations, the first excited state in the isospin-0 channel appears at a much lower energy than the first excited state in lattice QCD calculations. Our results also differ from results obtained for the linear sigma model by unitarization~\cite{Achasov_LSM_1994}.

The discrepancy with lattice QCD likely indicates the chiral effective expansion does not converge well with our lattice regularization.
\hl{However, since we have proved that an accurate lattice chiral effective action exists in \Cref{sec:exact-effective-theory-for-qcd}, we think this discrepancy}\sout{This}
might be alleviated in future works using smearing. \hl{If smearing does indeed solve the discrepancy with QCD, lattice ChEFT could provide useful information for lattice QCD extrapolations at the level of correlation functions, with the same finite volume treatment and in Euclidean time.
}

\section{Acknowledgments}

We thank our RBC and UKQCD collaborators, and especially Tom Blum, for helpful discussions and software support. L.J. and J.S. acknowledge the support of DOE Office of Science Early Career Award DE-SC0021147, DOE grants DE-SC0010339 and DE-SC0026314. We developed the computational code used for this work based on \href{https://github.com/jinluchang/Qlattice}{Qlattice} \cite{Qlattice}. The computational work for this project was conducted using resources provided by the Storrs High-Performance Computing (HPC) cluster. We extend our gratitude to the UConn Storrs HPC and its team for their resources and support, which aided in achieving these results.

\appendix

\section{Exact Effective Theory for QCD}
\label{sec:exact-effective-theory-for-qcd}
In this appendix, we demonstrate that it is possible to construct a lattice effective theory in terms of pions
which exactly reproduces the dynamics of lattice QCD.
Specifically, we construct a lattice field theory with four scalar fields whose correlation functions are exactly the same as the lattice QCD correlation functions of the corresponding quark bilinears.
This is an explicit construction of the effective theory action in terms of meson fields (baryons can also be included), a very different strategy compared to Ref.~\cite{Kawamoto:1981hw}.
For two flavor QCD, we can define
\ba
Q_0(x) = Z_\phi \bar\psi(x) \psi(x) \text{ and } Q_{i}(x) = Z_\phi \bar\psi(x) \tau_{i} i \gamma^5\psi(x)\text{ for }i=1,2,3,
\label{eq:quark-bilinear-op-def}
\ea
where $Z_\phi$ is an arbitrary normalization coefficient to match the quark bilinear operator with the desired scalar field.
The action of this effective theory can be explicitly constructed as
\ba
e^{-S_\text{eff}[\{\phi_i\}]}
=
\int \mathcal{D}U\mathcal{D}\bar\psi\mathcal{D}\psi
\exp\Big(-S_\text{QCD}[U, \bar \psi, \psi]-\frac{1}{2}\sum_{x}\sum_{i=0}^3 M_C^2\big(\phi_i(x)-Q_i(x)\big)^2\Big)
.
\label{eq:s-eff-def}
\ea
Note that the above integral may not always be positive, and so the resulting action $S_\text{eff}[\phi_i]$ is not guaranteed to be real. Nevertheless, if we calculate correlation functions $\mathcal{O}[\{\phi_i\}]$ with this action, we obtain 
\ba
\langle \mathcal{O}[\{\phi_i\}]\rangle
=&\frac{1}{Z}\int \mathcal{D}\phi\, e^{-S_\text{eff}[\{\phi_i\}]}\mathcal{O}[\{\phi_i\}]
\\
=&\frac{1}{Z}\int \mathcal{D}\phi \mathcal{D}U\mathcal{D}\bar\psi\mathcal{D}\psi
\nn\\&\times
\exp\Big(-S_\text{QCD}[U, \bar \psi, \psi]-\frac{1}{2}\sum_{x}\sum_{i=0}^3 M_C^2\big(\phi_i(x)-Q_i(x)\big)^2\Big)
\mathcal{O}[\{\phi_i\}]
\\
=&\frac{1}{Z}\int \mathcal{D}\phi' \mathcal{D}U\mathcal{D}\bar\psi\mathcal{D}\psi
\nn\\&\times
\exp\Big(-S_\text{QCD}[U, \bar \psi, \psi]-\frac{1}{2}\sum_{x}\sum_{i=0}^3 M_C^2{\phi_i'}^2(x)\Big)
\mathcal{O}[\{Q_i + \phi_i'\}]
\\
=&
\langle \mathcal{O}[\{Q_i + \phi_i'\}]\rangle_{\text{QCD},\phi'}
\ea
where
\ba
Z=&\int \mathcal{D}\phi\, e^{-S_\text{eff}[\{\phi_i\}]}
\\
=&
\int \mathcal{D}\phi' \mathcal{D}U\mathcal{D}\bar\psi\mathcal{D}\psi
\exp\Big(-S_\text{QCD}[U, \bar \psi, \psi]-\frac{1}{2}\sum_{x}\sum_{i=0}^3 M_C^2{\phi_i'}^2(x)\Big)
.
\ea
Thus we obtain a relationship between correlation functions in the effective scalar theory
and correlation functions in the theory of QCD plus a set of non-propagating fields with distribution $e^{-M_C^2{\phi_i'}^2/2}$
\ba
\langle \mathcal{O}[\{\phi_i\}]\rangle=\langle \mathcal{O}[\{Q_i + \phi_i'\}]\rangle_{\text{QCD},\phi'}.
\ea
The only effect these non-propagating fields will have on QCD correlation functions is to introduce some additional ``contact'' terms. For example,
\ba
\langle \phi_i(x) \phi_j(y) \rangle
=
\langle Q_i(x) Q_j(y) \rangle_{\text{QCD}}
+
\langle \phi_i'(x) \phi_j'(y) \rangle_{\phi'}
\ea
where
\ba
\langle \phi_i'(x) \phi_j'(y) \rangle_{\phi'}
=&
\frac{1}{M_C^2} \delta_{i,j} \delta^{(4)}(x-y)
.
\ea
In the limit of large $M_C$, even these additional contact terms vanish, and we have
\ba
\langle \mathcal{O}[\{\phi_i\}]\rangle=\langle \mathcal{O}[\{Q_i\}]\rangle_\text{QCD}\qquad(\text{when }M_C\to\infty).
\ea

Therefore, the effective theory described by $S_\text{eff}$ exactly reproduces the underlying QCD dynamics without any approximation. Also, it is straightforward to include additional fields (or exclude some of the existing fields) in the construction.

\subsection{Effective action in the chiral limit}

While this explicit construction does not immediately enable us to perform perturbative calculations or numerical simulations, we can prove a very simple relation between the effective action $S_\text{eff}$ and its form in the chiral limit $S_\text{eff}^{m=0}$.
Start with the relation of the QCD action:
\ba
S_\text{QCD}[U, \bar \psi, \psi] =& S_\text{QCD}^{m=0}[U, \bar \psi, \psi] + \sum_x m \bar \psi(x)\psi(x)
\\
=&
S_\text{QCD}^{m=0}[U, \bar \psi, \psi] + \sum_x \frac{m}{Z_\phi} Q_0(x)
\label{eq:qcd-relation}
\ea
where $S_\text{QCD}^{m=0}$ is the QCD action in the chiral limit and $m$ is the mass parameter for the up and down quarks.
Note that the $Q_0(x)$ is defined in Eq.~\eqref{eq:quark-bilinear-op-def}.
The chiral limit effective action $S_\text{eff}^{m=0}$ is constructed with the same approach as $S_\text{eff}$ in Eq.~\eqref{eq:s-eff-def}:
\ba
e^{-S_\text{eff}^{m=0}[\{\phi_i\}]}
=
\int \mathcal{D}U\mathcal{D}\bar\psi\mathcal{D}\psi
\exp\Big(-S_\text{QCD}^{m=0}[U, \bar \psi, \psi]-\frac{1}{2}\sum_{x}\sum_{i=0}^3 M_C^2\big(\phi_i(x)-Q_i(x)\big)^2\Big)
.
\label{eq:s-eff-def-chiral}
\ea
Based on \Cref{eq:s-eff-def,eq:qcd-relation,eq:s-eff-def-chiral},
we can express effective action $S_\text{eff}$ in terms of $S_\text{eff}^{m=0}$ 
\ba
e^{-S_\text{eff}[\{\phi_i\}]}
=&
\int \mathcal{D}U\mathcal{D}\bar\psi\mathcal{D}\psi
\exp\Big(-S_\text{QCD}[U, \bar \psi, \psi]-\sum_{x}\sum_{i=0}^3 \frac{1}{2} M_C^2\big(\phi_i(x)-Q_i(x)\big)^2\Big)
\\
=&
\int \mathcal{D}U\mathcal{D}\bar\psi\mathcal{D}\psi
\exp\Big(-S_\text{QCD}^{m=0}[U, \bar \psi, \psi]
\nn\\&\hspace{2cm}
-\sum_{x}
\Big(
\sum_{i=0}^3 \frac{1}{2} M_C^2\big(\phi_i(x)-Q_i(x)\big)^2
+
\frac{m}{Z_\phi} Q_0(x)
\Big)
\Big)
\\
=&
\int \mathcal{D}U\mathcal{D}\bar\psi\mathcal{D}\psi
\exp\Big(-S_\text{QCD}^{m=0}[U, \bar \psi, \psi]
\nn\\&\hspace{1cm}
-\sum_{x}
\Big(
\sum_{i=0}^3 \frac{1}{2} M_C^2\big(\phi_i(x)-\frac{m}{Z_\phi M_C^2}\delta_{i,0}-Q_i(x)\big)^2
\nn\\&\hspace{3cm}
+
\frac{m}{Z_\phi} \phi_0(x)
-
\frac{m^2}{2 Z_\phi^2 M_C^2}
\Big)
\Big)
\\
=&
\exp\Big(
-
S_\text{eff}^{m=0}[\{\phi_i-\frac{m}{Z_\phi M_C^2}\delta_{i,0}\}]
-
\sum_{x}
\frac{m}{Z_\phi} 
\phi_0(x)
+
\sum_{x}
\frac{m^2}{2 Z_\phi^2 M_C^2}
\Big)
.
\ea
Therefore, the general effective action can be expressed in terms of the chiral limit effective action
plus a simple linear term:
\ba
S_\text{eff}[\{\phi_i\}]
=&
S_\text{eff}^{m=0}[\{\phi_i-\frac{m}{Z_\phi M_C^2}\delta_{i,0}\}]
+
\sum_{x}
\frac{m}{Z_\phi} 
\phi_0(x)
-
\sum_{x}
\frac{m^2}{2 Z_\phi^2 M_C^2}
.
\ea

\subsection{Effective action with baryons}

We can include baryons with this style of explicit construction.
In the path integral, the fermion fields are represented by the anti-commuting Grassmann variables.
We start with the following identities:
\ba
\int d B \int d \eta\, e^{-\eta (f(\psi) - B)}
\,
&=
1
\\
\int d B \int d \eta\, e^{-\eta (f(\psi) - B)} B
\,
&=
f(\psi)
\ea
where $\eta, B, f(\psi)$ are Grassmann variables or functions of Grassmann variables. They anti-commute with themselves and with each other.
The identity generalizes to arbitrary functions of $B$ and to the multi-variable case:
\ba
\int d B \int d \eta\, e^{-\eta (f(\psi) - B)}  \mathcal O(B)
\,
&=
\mathcal O(f(\psi))
\\
\int \mathcal D B \int \mathcal D \eta\, 
\exp\big(-\sum_i \eta_i (f_i(\{\psi_j\}) - B_i)\big)
\mathcal O({B_i})
&=
\mathcal O({f_i(\{\psi_j\})})
\label{eq:baryon-eff-relation}
\ea

We can use these relations to formally construct an effective theory including both baryons and mesons.
For example, we can define the proton and neutron operators as
\ba
P(x)
=&
\epsilon_{a,b,c}
u_{a}(x) (u^T_b(x) C \gamma_5 d_c(x))
\\
N(x)
=&
\epsilon_{a,b,c}
d_{a}(x) (d^T_b(x) C \gamma_5 u_c(x))
\\
\bar P(x)
=&
-\epsilon_{a,b,c}
\bar u_{a}(x) (\bar u_b(x) C \gamma_5 \bar d^T_c(x))
\\
\bar N(x)
=&
-\epsilon_{a,b,c}
\bar d_{a}(x) (\bar d_b(x) C \gamma_5 \bar u^T_c(x))
\ea
where $u$, $d$, $\bar u$, $\bar d$, and $C$ can be defined as
\ba
\psi(x) =& \left(\begin{array}{c} u(x) \\ d(x)\end{array} \right)
\\
\bar\psi(x) =& \left(\begin{array}{cc} \bar u(x) & \bar d(x) \end{array}\right)
\\
C=& \gamma_t\gamma_y
\ea
The effective action can be constructed as
\ba
e^{-S_\text{eff}[\{\phi_i\},\bar p,p,\bar n,n]}
=&
\int \mathcal{D}U\mathcal{D}\bar\psi\mathcal{D}\psi
\mathcal{D}\bar \eta_p
\mathcal{D}\eta_p
\mathcal{D}\bar \eta_n
\mathcal{D}\eta_n
\nn\\&\hspace{-2cm}\times
\exp\Big(
-S_\text{QCD}[U, \bar \psi, \psi]
-\frac{1}{2}\sum_{x}\sum_{i=0}^3 M_C^2\big(\phi_i(x)-Q_i(x)\big)^2
\nn\\&\hspace{-1cm}
-
\sum_{x} (\bar p(x) - \bar P(x)) \eta_p(x)
-
\sum_{x} \bar \eta_p(x) (p(x) - P(x))
\nn\\&\hspace{-1cm}
-
\sum_{x} (\bar n(x) - \bar N(x)) \eta_n(x)
-
\sum_{x} \bar \eta_n(x) (n(x) - N(x))
\Big)
.
\label{eq:s-eff-def-b}
\ea
Based on the previous derivation in Eq.~\eqref{eq:baryon-eff-relation}, this construction satisfies the following desired property
\ba
&\langle \mathcal{O}[\{\phi_i\},\bar p,p,\bar n,n] \rangle
\nn\\
=&\frac{1}{Z}\int \mathcal{D}\phi
\mathcal{D}\bar p
\mathcal{D}p
\mathcal{D}\bar n
\mathcal{D}n
\, e^{-S_\text{eff}[\{\phi_i\},\bar p,p,\bar n,n]}
\mathcal{O}[\{\phi_i\},\bar p,p,\bar n,n]
\\
=&\frac{1}{Z}\int \mathcal{D}\phi' \mathcal{D}U\mathcal{D}\bar\psi\mathcal{D}\psi\,
\exp\Big(-S_\text{QCD}[U, \bar \psi, \psi]-\frac{1}{2}\sum_{x}\sum_{i=0}^3 M_C^2{\phi_i'}^2(x)\Big)
\nn\\&\times
\mathcal{O}[\{Q_i + \phi_i'\},\bar P,P,\bar N,N]
\\
=&
\langle
\mathcal{O}[\{Q_i + \phi_i'\},\bar P,P,\bar N,N]
\rangle_{\text{QCD},\phi'}
\ea
where
\ba
Z=&\int \mathcal{D}\phi
\mathcal{D}\bar p
\mathcal{D}p
\mathcal{D}\bar n
\mathcal{D}n
\, e^{-S_\text{eff}[\{\phi_i\},\bar p,p,\bar n,n]}
\\
=&
\int \mathcal{D}\phi' \mathcal{D}U\mathcal{D}\bar\psi\mathcal{D}\psi
\exp\Big(-S_\text{QCD}[U, \bar \psi, \psi]-\frac{1}{2}\sum_{x}\sum_{i=0}^3 M_C^2{\phi_i'}^2(x)\Big)
.
\ea

This concludes our explicit and exact construction of the effective theory.

\hl{
\section{Correlation Functions}\label{sec:correlation_functions}
We use the following correlation functions to define observables:
\begin{itemize}
    \item To determine the pion mass, we define $\mathcal{O}_\pi(t)\equiv \sum_{i=1}^3\sum_{\mathbf{x}}\phi_i(\mathbf{x},t)$ and fit 
    \ba
    \left\langle T\left[\sum_t\mathcal{O}_\pi(t+\Delta t)\mathcal{O}_\pi(t)\right]\right\rangle = A \cosh\left[\left(\frac{N_t}{2}-\Delta t\right) m_\pi\right],
    \ea 
    where $N_t$ is the number of lattice sites in the time direction. We use $\cosh[(N_t/2-t)m_\pi]$ instead of $\exp(-tm_\pi)$ because of the periodic boundary conditions.
    \item For the effective sigma mass, we define $\mathcal{O}_\sigma(t)\equiv -\sum_{\mathbf{x}}\phi_0(\mathbf{x},t)$ and take $m_\sigma^\text{eff}=-\ln[C_\sigma(1)/C_\sigma(0)],$
    where
    \ba\nn
    C_\sigma(\Delta t) = \left\langle T\left\{\sum_t\left[\mathcal{O}_\sigma(t+\Delta t)-\langle\mathcal{O}_\sigma\rangle\right]\left[\mathcal{O}_\sigma(t)-\langle\mathcal{O}_\sigma\rangle\right]\right\}\right\rangle
    \ea
    and $\langle\mathcal{O}_\sigma\rangle$ is the vacuum expectation value of $\mathcal{O}_\sigma$.
    \item To determine $F_\pi$, we define the axial current operator $A^i_\mu(x)=\phi_0(x-\mu)\phi_i(x)-\phi_0(x)\phi_i(x-\mu)$, as discussed in \Cref{sec_pion_decay_constant}. On a finite lattice, the right-hand side of \Cref{eq:Fpi-definition} will become
    \ba
    F_\pi m_\pi e^{-m_\pi t} \to 2F_\pi m_\pi e^{-m_\pi N_t/2}\sinh\left[\left(\frac{N_t}{2}-t+\frac{1}{2}\right)m_\pi\right].
    \ea
    We use $t-1/2$ instead of $t$ in the above equation because the lattice definition of $A_0^i$ combines fields at $t$ and $t-1$.
    To create the pion state, we define the pion interpolating operator
    \ba 
    \mathcal{O}^i_\pi(t)\equiv \sum_\mathbf{x} \phi^i(\mathbf{x},t).
    \ea 
    Taking into account the correct normalization of the pion state, we get
    \ba \nn
    \left\langle T\left\{\left(\frac{1}{L^3}\sum_\mathbf{x}A_0^i(\mathbf{x},t)\right)\mathcal{O}^i_\pi(0)\right\}\right\rangle\sqrt{2m_\pi L^3}\sqrt\frac{2e^{-m_\pi N_t/2}\cosh((N_t/2-t)m_\pi)}{\langle0| \mathcal{O}_\pi^i(t)\mathcal{O}_\pi^i(0)|0\rangle}
    \ea 
    \ba \label{eq:Fpi-fit}
    = 2F_\pi m_\pi e^{-m_\pi N_t/2}\sinh\left[\left(\frac{N_t}{2}-t+\frac{1}{2}\right)m_\pi\right]
    \ea
    We can extract the decay constant using \Cref{eq:Fpi-fit}, but to get a more stable fit, we can instead use Noether's theorem for the lattice to get $\sum_\mu \big(A_\mu^i(x+\mu)-A_\mu^i(x)\big)=-\alpha\phi^i(x).$ To get our results, we combined this relationship with \Cref{eq:Fpi-fit} to get a correlation function we could fit in order to extract $F_\pi$.
\end{itemize}
}

\bibliography{linear_sigma_model}

\end{document}